\documentclass[twocolumn,prl,noshowpacs,superscriptaddress]{revtex4-1}
\usepackage[latin9]{inputenc}
\usepackage{float}
\usepackage{bm}
\usepackage{amsmath}
\usepackage{amssymb}
\usepackage{graphicx}

\makeatletter
\usepackage{graphics}
\usepackage{bm}
\usepackage{xcolor}
\usepackage{dcolumn}
\usepackage{bm}
\newcommand{\bq}{\begin{equation}}
\newcommand{\eq}{\end{equation}}

\newcommand{\bfk}{{\bf k}}
\usepackage{epstopdf}
\newcommand{\yrs}{YbRh$_2$Si$_2$\,}
\newcommand{\cca}{CeCu$_{1-x}$Au$_x$\,}

\makeatother

\begin{document}

\title{Superconductivity at an antiferromagnetic quantum critical point:
the role of energy fluctuations}

\author{Jian Kang}

\affiliation{National High Magnetic Field Laboratory, Tallahassee, FL 32304}

\author{Rafael M. Fernandes}

\affiliation{School of Physics and Astronomy, University of Minnesota, Minneapolis
MN 55455}

\author{Elihu Abrahams} \thanks{Deceased October 18, 2018}

\affiliation{Department of Physics and Astronomy, University of California Los
Angeles, Los Angeles, CA 90095}

\author{Peter W\"olfle}

\affiliation{Institute for Theory of Condensed Matter, Karlsruhe Institute of
Technology, 76049 Karlsruhe, Germany}

\affiliation{Institute for Nanotechnology, Karlsruhe Institute of Technology,
76031 Karlsruhe, Germany}
\begin{abstract}
Motivated by recent experiments reporting superconductivity only at very
low temperature in a class of heavy fermion compounds, we study the impact of
energy fluctuations with small momentum transfer on the pairing instability
near an antiferromagnetic quantum critical point. While these fluctuations,
formed by composite spin fluctuations, were proposed to explain the 
thermodynamic and transport properties near the quantum critical point of 
compounds such as YbRh$_{2}$Si$_{2}$ and CeCu$_{6-x}$Au$_{x}$ at 
$x\approx0.1$, here they are found to strongly suppress $T_{c}$ of 
the $d$-wave pairing of the hot quasiparticles promoted
by the spin fluctuations. Interestingly, if energy fluctuations are strong enough, 
they can induce triplet pairing involving the quasiparticles of the cold regions 
of the Fermi surface. Overall, the opposing effects of energy and spin fluctuations
lead to a suppression of $T_{c}$.
\end{abstract}
\maketitle

\section{Introduction}

One of the interesting issues associated with a magnetic quantum critical
point (QCP) is the possibility of superconductivity induced by the
coupling between the associated quantum critical fluctuations and
the electron quasiparticles \cite{AVC03,Sachdev10,Zaanen11,Berg12,Senthil15,Raghu15,SSLee15,Wang16}.
There are a number of heavy-fermion compounds \cite{1} that exhibit
antiferromagnetic quantum criticality and superconductivity nearby
in their phase diagram. Superconductivity in the cuprate \cite{Scalapino12}
and iron-based \cite{Hirschfeld11,AVC12} compounds is often argued
to be a consequence of the presence of strong magnetic fluctuations
. However, there are some prominent cases of heavy-fermion antiferromagnetic
quantum criticality in which nearby superconductivity is either absent
(\cca) or has a tiny transition temperature $T_{c}$, if at all (\yrs)
\cite{schu}. Elucidating why superconductivity is absent (or so fragile)
in these cases, despite the presumed presence of strong magnetic fluctuations,
is an important issue in the field of unconventional superconductivity

Here, we address this issue in the framework of the recently-developed
theory of critical quasiparticles whose properties are generated by
their interaction with critical antiferromagnetic fluctuations \cite{qp1,exp1,WSA}.
One of the outcomes of this model is the importance of low-energy,
small-momentum composite spin fluctuations, dubbed energy fluctuations
\cite{exp2,ward}. Previously, it was shown that these energy fluctuations
can explain unusual thermodynamic and transport properties observed in certain heavy
fermion compounds near their magnetic QCP. In this paper, we apply an Eliashberg-like
approach to investigate the interplay between spin and energy critical
fluctuations to the pairing problem in a three-dimensional system.

We find that the contribution of each fluctuation channel depends
strongly on the quasiparticle position on the Fermi surface (FS).
It is well-known that antiferromagnetic (AFM) spin fluctuations with
wave-vector ${\bf Q}$ pair quasiparticles in the ``hot line\char`\"{}
regions of the FS, i.e. the regions for which the quasiparticle energies
$\epsilon_{{\bf k}}$ and $\epsilon_{{\bf k}+{\bf Q}}$ are equal
\cite{Varma86,Scalapino86,Millis88,Pines92}. The quasiparticles in
the remaining ``cold\char`\"{} parts of the FS are little affected.
Thus, single spin fluctuation exchange can be attractive for hot quasiparticles
and results in a non-zero $T_{c}$ for $d$-wave singlet superconductivity.
However, we find that the exchange of energy fluctuations is in general
repulsive in that channel and may substantially reduce $T_{c}$, even
to zero. On the other hand, exchange of energy fluctuations between
cold quasiparticles may induce spin-triplet $p$-wave superconductivity,
if only at a substantially lower temperature.

The paper is organized as follows: Section II reviews the strong-coupling
theory of critical quasiparticles, and the emergence of energy fluctuations.
Section III establishes the Eliashberg-like equations to study pairing
mediated by both spin and density fluctuations. These equations are
then solved in Section IV in both singlet and triplet channels. Section
V is devoted to the conclusions.

\section{Critical Quasiparticles: Normal state properties}

In this section, we briefly outline the main results of the theoretical
approach introduced in \cite{qp1,WSA}. The usual approach for heavy-fermion
metals that exhibit an antiferromagnetic quantum critical point involves
consideration of the interaction of fermionic quasiparticles with
the bosonic critical spin fluctuations. This may cause the fermionic
degrees of freedom to also have critical behavior that acts back on
the boson spectrum. This was first analyzed self-consistently in the
theory of critical quasiparticles \cite{qp1,WSA}, which was found
to have two qualitatively different solutions, one in the weak-coupling
and the other in the strong-coupling regime. The strong-coupling regime
gives the power laws that govern transport and thermodynamic properties
in the neighborhood of the QCP; it successfully accounts for experimental
results in both \yrs \cite{exp1} and \cca \cite{exp2}. In particular,
it was found that the quasiparticle weight factor $Z(\omega,T)\propto\lbrack\max(\omega,T)]^{\eta}\to0$
has a dimension-dependent fractional power of $\max(\omega,T)$. The
exponents $\eta$ on the cold and hot parts of the Fermi surface in
the case of three-dimensional spin fluctuations were found to be $\eta_{c}=1/4$\ and
$\eta_{h}=1/2$, respectively. This leads in turn to singular critical
behavior of various interaction vertex functions that are related
to $Z^{-1}$ by Ward identities \cite{ward}.

\begin{figure}[h]
{\centering \hspace{1.2in} \includegraphics[width=0.36\textwidth]{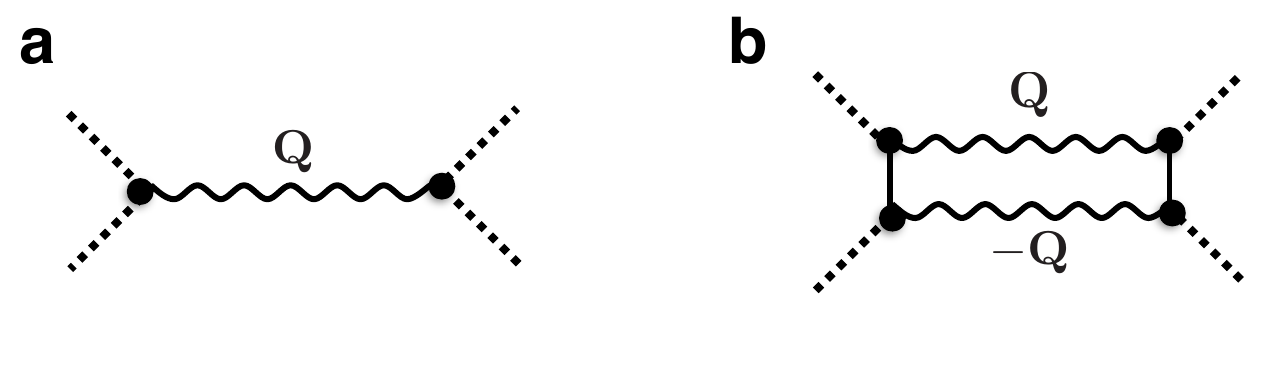}}
\caption{Critical fluctuations: \textbf{a}.\ Single spin fluctuation $\chi$
peaked at the AFM ordering vector ${\bf Q}$. \textbf{b}.\ Structure
of the energy fluctuation $\chi_{E}$. A second contribution has the
two spin fluctuation lines crossed.The dashed lines represent the
particle and hole excitations at the Fermi surface to which the fluctuations
couple. The full lines are excitations far from the Fermi surface,
and the black dots represent the vertex function $\Lambda_{Q}$.}
\label{F1}
\end{figure}

The typical antiferromagnetic ordered phase is usually characterized
by an ordering wave vector ${\bf Q}$. As discussed above, the associated
critical fluctuations then connect the special hot-spot regions of
the FS, which follow the condition $\epsilon_{\mathbf{k}}=\epsilon_{\mathbf{k}+\mathbf{Q}}$,
where $\epsilon_{\mathbf{k}}$ is the single-electron dispersion.
In three dimensional FS, this gives rise to hot lines. As a consequence,
the quasiparticle self energy generated by the exchange of such fluctuations
is highly anisotropic and critical mainly at the hot spots. However,
the exchange of \textit{two} spin fluctuations with total momentum
near zero \cite{hart}, which may be viewed \cite{WSA} as a spin
exchange-energy fluctuation, gives a critical contribution over the
whole FS (see Fig. \ref{F1}). The critical enhancements of the interaction
vertices mentioned above make such energy fluctuations important near
the QCP, both for their effect on the quasiparticle self energy and
for their role in superconductive pairing.

The spectrum of critical spin fluctuations is determined by the dynamical
spin susceptibility
\begin{equation}
\chi({\bf q},\nu)=\frac{N_{0}}{r+({\bf q-Q})^{2}\xi_{0}^{2}-i\Lambda_{Q}^{2}(\nu/v_{F}Q)}\label{X}
\end{equation}
where $N_{0}$ is the bare density of states at the Fermi level, $r$
is the control parameter tuning the system through the QCP, $\xi_{0}\approx k_{F}^{-1}$
is a microscopic correlation length, ${\bf Q}$ is the AFM ordering
vector, $v_{F}=k_{F}/m$ is the bare Fermi velocity, and $\Lambda_{Q}=\Lambda(\mathbf{k},\omega=0;\mathbf{q},\nu)$
is the vertex function for the antiferromagnetic spin fluctuation-particle-hole
interaction, \textit{i.e.} the vertex at frequency transfer $\nu$
and non-zero momentum transfer ${\bf q}\approx{\bf Q}$. Its presence
in the Landau damping term of Eq.\ ({\ref{X}}) reflects the feedback
into the critical bosonic spin fluctuations by the critical behavior
of the quasiparticles. It may be shown that when $Z^{-1}(\omega)$
diverges, then the vertex $\Lambda_{Q}\sim Z^{-1}$ will diverge as
well. For three-dimensional spin fluctuations, $\Lambda_{Q}(\nu)\propto\nu^{-\eta_{c}}$
\cite{ward}. We note here that the static susceptibility $\chi(r,\mathbf{q},\nu)$
diverges at $r=0$, $\mathbf{q}=\mathbf{Q}$, and $\nu=0$. However,
at non-zero temperature, $r$ does not diverge, i.e. the correlation
length is finite, following $r\sim T^{1-2\eta}$.

We define the energy fluctuation propagator $\chi_{E}({\bf q},\nu)$
as the composite of two spin fluctuations with total momentum ${\bf q}$
near zero. The relevant diagram is shown in Fig.\ \ref{F1}b. Schematically,
$\chi_{E}({\bf q},\nu)\sim\sum_{q_{1},\nu_{1}}G\cdot G\cdot\chi(q_{1},\nu_{1})\cdot\chi(q_{1}-q,\nu_{1}-\nu)$,
where one $\chi$ is peaked near ${\bf Q}$, the other near $-{\bf Q}$.
The two fermion propagators $G$, represented by the vertical lines
in the figure are both far from the FS, when the fluctuation couples
to particle and hole excitations (represented by dashed lines) near
the FS. The calculation, including both parallel and crossed contributions
to Fig.\ (1b) \cite{WSA} yields
\begin{equation}
\mathrm{Im}\,\chi_{E}(\mathbf{q},i\nu_{n})\approx N_{0}^{3}\Lambda_{Q}^{2d-3}\frac{|\nu_{n}/\gamma|^{d-1/2}}{\left[r+q^{2}\xi_{0}^{2}+|\nu_{n}|\Lambda_{Q}^{2}/\gamma\right]^{(d+1)/2}},
\end{equation}
where $\gamma$ is an energy scale of order the Fermi energy (\textit{e.g.}
$v_{F}Q$) and $d$ is the dimensionality of the spin fluctuations
\cite{dim}. In $d=3$ dimensions, and on the imaginary frequency
axis, the dependence of $\chi_{E}(\mathbf{q},i\nu_{n})$\ on $\mathbf{q},i\nu_{n}$
is similar to that of $\chi(\mathbf{q},i\nu_{n})$, except that $\chi_{E}$
diverges at $\mathbf{q}=0$. That is,
\begin{equation}
\chi_{E}(\mathbf{q},i\nu_{n})\approx N_{0}^{2}\Lambda_{Q}|\nu_{n}/\gamma|^{3/2}\chi(\mathbf{q+Q},i\nu_{n}).\label{XE}
\end{equation}

The role of both $\chi$ and $\chi_{E}$ on the normal-state properties
of the heavy fermion compounds has been investigated in Refs. \cite{exp2,WSA}.
Our goal here is to assess their interplay for the pairing instability
that arises near the antiferromagnetic QCP.

\section{Eliashberg Equations: Superconducting state properties}

To analyze the contributions of the critical fluctuations to pairing,
we consider the Eliashberg-like gap equation:
\begin{equation}
\Phi_{\alpha\beta}(\mathbf{k},i\omega_{n})=-T\sum_{\omega_{m}\mathbf{p},\gamma\delta}\frac{V_{\alpha\beta,\gamma\delta}(\mathbf{k}-\mathbf{p},i\omega_{nm})\Phi_{\gamma\delta}(\mathbf{p},i\omega_{m})}{\omega_{m}^{2}Z_{m}^{-2}+\epsilon_{\mathbf{p}}^{2}+|\Phi(\mathbf{p},i\omega_{m})|^{2}},
\end{equation}
where $Z_{m}^{-1}=1-\Sigma(i\omega_{m})/i\omega_{m}$ is the quasiparticle
weight factor determined by the ``second\char`\"{} Eliashberg equation.
In this work, we will not solve the second Eliashberg equation, and
instead will use the previously published results for the frequency
dependence of $Z$ in the strong-coupling regime of the model discussed
above \cite{qp1,WSA}. Here, $\omega_{n},\omega_{m}$ and $\omega_{nm}=\omega_{n}-\omega_{m}$\ are
fermionic and bosonic Matsubara frequencies, $\alpha,\beta,\gamma,\delta$
\ are spin indices and the summation over momentum $\mathbf{p}$
extends over the first Brillouin zone. As we shall only discuss the
superconducting $T_{c}$, we may drop $|\Phi|^{2}$ in the denominator
(``linearized gap equation\char`\"{}) and eventually take $Z$ to
be the normal state quasiparticle weight. As mentioned above, $Z$
has been calculated in Ref \onlinecite{WSA} as $Z=(\omega/E_{F})^{\eta}$,
where $\eta_{c}=1/4$ on the cold part of the Fermi surface and $\eta_{h}=1/2$
at the hot spots.

The pairing interaction $V(\bfk-{\bf q},i\omega_{n}-i\omega_{m})$
has two contributions: one from the exchange of a single spin fluctuation,
Eq.\ ({\ref{X}}), the other from exchange of an energy fluctuation,
Eq.\ (\ref{XE}). Both interactions are of the spin exchange type,
\begin{align}
V_{\alpha\beta,\gamma\delta} & =V{\bm{\tau}}_{\alpha\gamma}\cdot{\bm{\tau}}_{\beta\delta}\nonumber \\
 & =V_{s}(i\tau_{\alpha\beta}^{y})(i\tau_{\gamma\delta}^{y})+V_{t}(i\tau^{y}{\bm{\tau}})_{\alpha\beta}\cdot(i\tau^{y}{\bm{\tau}})_{\gamma\delta},
\end{align}
where ${\bm{\tau}}=(\tau^{x},\tau^{y},\tau^{z})$ is the vector of
Pauli matrices. The last equation displays the spin dependence in
the particle-particle channel. The singlet and the triplet parts are
given by $V_{s}=3V$ and $V_{t}=-V$, where

\begin{equation}
V(\mathbf{q},i\nu_{n})=\alpha^{2}\chi(\mathbf{q},i\nu_{n})+4h(\nu_{n})\alpha_{E}^{2}\chi_{E}(\mathbf{q},i\nu_{n}).\label{V_general}
\end{equation}

Here, we shall approximate the coupling constants $\alpha$ as $\alpha \approx \Lambda_{Q}/N_{0}$
and $\alpha_{E}\approx\Lambda_{v}(\Lambda_{Q}/N_{0})^{2}$.
The vertex function $\Lambda_{Q}$ at each end of a spin fluctuation
was introduced below Eq.\ ({\ref{X}}) and $\Lambda_{v}\approx Z^{-1}$
is the vertex at each end of an energy fluctuation. We have introduced
the function $h(\nu_{n})=[\exp(5(|\nu_{n}|/\nu_{c}-1))+1]^{-1}$ which
gives a soft cutoff at $\nu_{c}\ll\epsilon_{F}$ for the energy fluctuations.  As for the spin fluctuations, we include the hard cutoff $\Lambda_{cut} = \epsilon_f$.

As argued in Ref.\ \onlinecite{exp2}, for a quantum critical system
to enter the strong coupling regime, as we have assumed, it is necessary
that some additional quantum fluctuations, such as ferromagnetic fluctuations,
should increase $Z^{-1}$ sufficiently and actually dominate the AFM
spin energy contributions when $\nu_{n}>\nu_{c}$. In the case of
\yrs, the crossover from the low temperature regime, characterized
by power-law behavior (\textit{e.g.} specific heat coefficient $C/T\propto T^{-\eta_{c}}$
to the high-$T$ behavior $C/T\propto\ln(T_{0}/T)$) occurs at $T\approx0.3$
K. If we take the characteristic Fermi temperature at $10$ K, we
deduce a frequency cutoff $\nu_{c}\approx0.03\epsilon_{F}$.

Although the singlet interaction is repulsive ($V_{s}>0$), as is
well-known, the exchange of a single AF spin fluctuation that is peaked
at ${\bf Q}$ connects quasiparticles at hot spots ${\bf k}_{h}$
and ${\bf k}_{h}+{\bf Q}$, which are usually far apart on the FS.
This mechanism often leads to unconventional pairing of quasiparticles
at the hot regions of the FS, characterized by a gap function $\Phi$
whose sign changes between these two hot spots (as would be the case
for a suitable $d$-wave gap symmetry)\cite{moriya}. Since cold quasiparticles
are boosted off the FS by scattering from a single spin fluctuation,
the cold regions do not contribute substantially to pairing via single
spin fluctuation exchange in our scenario (see also Ref.~\cite{Xiaoyu17}).
It will be seen that exchange of energy fluctuations (peaked at ${\bf q}\sim0$)
gives a {\em repulsive} contribution to the pairing kernel, as
it connects $\bfk_{h}+{\bf q}\approx\bfk_{h}$ for which the gap function
has the {\em same} sign. Therefore, we investigate below the suppression,
by energy fluctuations, of $d$-wave singlet superconductivity from
the hot regions.

As well as being repulsive in the singlet channel, the exchange of
energy fluctuations in the triplet channel will be attractive provided
it couples close regions of the FS (as it does, since $q\approx0$)
for which the gap function does \textit{not} change sign
\begin{equation}
V_{t}=-4h(\nu_{n})\alpha_{E}^{2}\chi_{E}({\bf q},i\nu_{n}).\label{trip}
\end{equation}
This pairing interaction is equally strong over the whole FS and so
could lead to triplet pairing of cold quasiparticles. The orbital
symmetry of the resulting gap function will likely be the most symmetric
form compatible with the requirement of odd-parity imposed by the
Pauli principle, \textit{e.g.} $p$-wave pairing in the present case.

\section{Calculation of $T_{c}$}

\begin{figure}[h]
{\centering \hspace{1.2in} \includegraphics[width=0.35\textwidth]{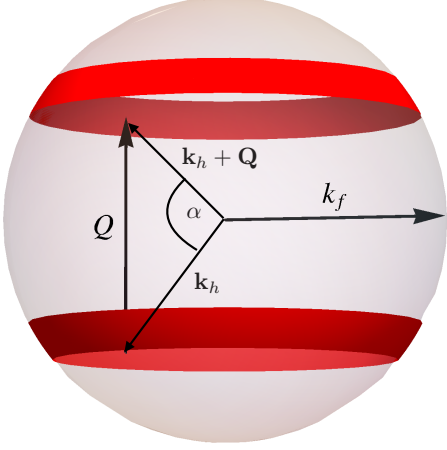}}
\caption{Schematic plot of the hot lines (red) on the Fermi surface. The hot
lines satisfy the condition $\epsilon_{\bfk_{h}}=\epsilon_{\bfk_{h}+{\bf Q}}=0$,
where $\mathbf{Q}$ is the antiferromagnetic wave-vector. Note that
the hot lines have a finite width. }
\label{F2}
\end{figure}

For the actual solution of the linearized gap equation, we take a
simple isotropic model of a three-dimensional metal with dispersion
$\epsilon_{\bfk}\approx v_{F}(k-k_{F})$ and three-dimensional antiferromagnetic
fluctuations as is appropiate for \yrs. The spherical FS has lines
of hot spots $\bfk_{h}$, where $\epsilon_{\bfk_{h}}=\epsilon_{\bfk_{h}+{\bf Q}}=0$.
Fig.\ \ref{F2} shows the two hot lines on the FS (in red) that are
connected by the AFM vector ${\bf Q}$ taken here to be parallel to
the $z$ axis. The hot lines are located at polar angle $\theta_{0}=\cos^{-1}(Q/2k_{F})$
and at $\pi-\theta_{0}$. The width of the hot lines \cite{WSA} depends
on the temperature as $\delta\theta\approx\Lambda_{Q}\sqrt{T/\epsilon_{F}}\sin\alpha$,
where $\alpha=\pi-2\theta_{0}$ is the angle between the quasiparticle
velocities ${\bf v}_{{\bfk_{h}}}$ and ${\bf v}_{{\bfk_{h}+{\bf Q}}},$
see Fig.~\ref{F2}.

\subsection{Hot Quasiparticles}

As explained earlier, we will restrict the analysis of singlet pairing
to the neighborhood of the hot lines. The linearized gap equation
has the form
\begin{equation}
\Phi(\mathbf{k},i\omega_{n})=-3T\sum_{\omega_{m};\mathbf{p;}\gamma,\delta}\frac{V(\mathbf{k}-\mathbf{p},i\omega_{nm})\Phi(\mathbf{p},i\omega_{m})}{\omega_{m}^{2}Z_{m}^{-2}+\epsilon_{\mathbf{p}}^{2}},
\end{equation}
where
\begin{align}
V(\mathbf{q},i\nu_{n})= & \alpha^{2}\chi(\mathbf{q},i\nu_{n})+\nonumber \\
 & 4h(\nu_{n})\alpha_{E}^{2}N_{0}^{2}\Lambda_{Q}|\nu_{n}|^{3/2}\chi(\mathbf{q+Q},i\nu_{n})\ ,\label{comb}
\end{align}
where the second term comes from $\chi_{E}$ in Eq.\ (\ref{XE}).
Neglecting the dependence of the gap function on $|\mathbf{p}|$,
its dependence is only on the polar angle $\theta$ and the azimuthal
angle $\phi$ since $\mathbf{p}$ is on the FS, \textit{i.e.} $|\mathbf{p}|=k_{F}$.
On the one hand, the Pauli principle requires the gap function on
the hot lines ($\theta=\theta_{0}$ and $\theta=\pi-\theta_{0}$)
to obey $\Phi(\theta_{0},\phi,i\omega_{m})=$ $\Phi(\pi-\theta_{0},\pi+\phi,i\omega_{m})$.
On the other hand, as explained above, the gap must change sign between
the two hot lines in order to solve the gap equation, $\Phi(\mathbf{p},i\omega_{m})\approx-\Phi(\mathbf{p+Q},i\omega_{m})$.
Combining these two conditions yields $\Phi(\theta_{0},\phi,i\omega_{m})=-\Phi(\theta_{0},\pi+\phi,i\omega_{m})$.
We therefore look for a solution of the form $\Phi(\theta,\phi,i\omega_{m})=\Delta_{m}^{s}\cos\theta\cos\phi$
defined along the hot lines.

In the first term of the effective interaction in Eq.\ (\ref{comb}),
we may shift $\mathbf{p\rightarrow p-Q}$, which leaves the factor
$(\omega_{m}^{2}Z_{m}^{-2}+\epsilon_{\mathbf{p}}^{2})^{-1}$ invariant,
since $\epsilon_{\mathbf{p-Q}}=\epsilon_{\mathbf{p}}$\ on the hot
lines, while the sign of $\Phi$ changes. We choose the proper sign
of ${\bf Q}$, depending on whether ${\bf p}$ is on the upper or
lower hot line (see Fig.\ \ref{F2}). Therefore, the interaction
function simplifies to $V(\mathbf{q},i\nu_{n})\rightarrow\lbrack-\alpha^{2} + 4 h_{nm} \alpha_{E}^{2}N_{0}^{2}\Lambda_{Q}|\nu/\gamma|^{3/2}]\chi(\mathbf{q+Q},i\nu_{n})$.

The linearized gap equation then takes the form
\begin{align}
 & \Delta_{n}^{s} \cos\theta_{k}\cos\phi_{k} \nonumber \\
= & 3T\sum_{\omega_{m};\mathbf{p}}\frac{N_{0}}{r+(\mathbf{k}-\mathbf{p})^{2}/ k_F^2 + \Lambda_{Q,nm}^{2}|\omega_{nm}|}\nonumber \\
 & \times\lbrack\alpha_{nm}^{2}-4h_{nm}\alpha_{E;nm}^{2}N_{0}^{2}\Lambda_{Q,nm}|\omega_{nm}|^{3/2}]\nonumber \\
 & \times\frac{\Delta_{m}^{s}\cos\theta_{p}\cos\phi_{p}}{\omega_{m}^{2}Z_{m}^{-2}+\epsilon_{\mathbf{p}}^{2}},\label{Eq:HotGapEqn}
\end{align}

Here, the momentum integration is restricted to the hot lines. For
not too large $|\omega_{m}|\ll\epsilon_{F}$ the factor $[\omega_{m}^{2}Z_{m}^{-2}+\epsilon_{\mathbf{p}}^{2}]^{-1}$
is sharply peaked at $p=k_{F}$, so that one may write $(\mathbf{k}-\mathbf{p})^{2}=2 k_F^2 (1-\cos\theta_{kp})$,
where $\theta_{kp}$ is the angle enclosed by $(\mathbf{k,p})$. If
we take $\mathbf{k} = k_F (\sin\theta_{0},0,-\cos\theta_{0}) $ on the lower
hot line, we have $(\mathbf{k}-\mathbf{p})^{2}=2 k_F^2 [1+\cos\theta_{0}\cos\theta_{p}-\sin\theta_{0}\sin\theta_{p}\cos\phi_{p}]$.

The integration over the angles $\phi_{p}$ and $\theta_{p}$ as well
as the integration over $\epsilon_{{\bf p}}$ can all be done analytically.
This results in a matrix equation in frequency space:
\begin{equation}
\Delta_{n}^{s}=\frac{3\pi T_{c}}{2}\sum_{m}W_{nm}^{s}\frac{\Delta_{m}^{s}}{|\omega_{m}|/Z_{m}},
\end{equation}

where

\begin{equation}
W_{nm}^{s}=[\Lambda_{c,nm}^{2}-4h_{nm}\Lambda_{c,nm}^{5}\Lambda_{h,nm}^{2}|\Omega_{nm}|^{3/2}]I_{nm}^{s}
\end{equation}
with

\begin{eqnarray}
I_{nm}^{s} & = & \frac{1+B_{nm}^{s}}{\sqrt{1+B_{nm}^{s}/2}}\sinh^{-1}\frac{\delta\theta}{\sqrt{2\sin^{2}\theta_{0}B_{nm}^{s}}}-\frac{\delta\theta}{\sin\theta_{0}},\nonumber \\
B_{nm}^{s} & = & \Lambda_{c,nm}^{2}|\Omega_{nm}|/2\sin^{2}\theta_{0}
\end{eqnarray}
and $h_{nm}=h(\omega_{n}-\omega_{m})$\ is the soft cutoff function
introduced earlier. It is convenient to define $f_{m}^{s}=\Delta_{m}^{s}Z_{m}/|\omega_{m}|$\ and
to re-express the gap equation as the matrix eigenvalue equation \cite{Kang16}
\begin{equation}
\sum_{m}K_{n,m}^{s}f_{m}^{s}=0, \label{Eq:HotGapMatrix}
\end{equation}
where the kernel is given by
\begin{align}
K_{n\neq m}^{s} & =\frac{3}{2}W_{nm}^{s},\nonumber \\
K_{n,n}^{s} & =-(2n+1)Z_{n}^{-1}+\frac{1}{2}(K_{n,n-1}^{s}+K_{n-1,n}^{s})\label{hotgap}
\end{align}

Here, the subscript $nm$ stands for the frequency difference $\omega_{n}-\omega_{m}$.
We have regularized the weak singularity of $K_{n,m}^{s}$ in the
limit $n\rightarrow m$, which is cutoff by temperature as noted in
the text below Eq.\ ({\ref{X}}), by setting $K_{n,n}^{s}\approx\frac{1}{2}(K_{n,n-1}^{s}+K_{n-1,n}^{s})$. The
subscripts $c,h$ label cold or hot quasiparticle quantities. Following
Ref. \cite{WSA}, we set $\Lambda_{Q}=\Lambda_{c,nm}=Z_{c,nm}^{-1}=1+\Lambda_{c}^{(0)}|\omega_{n}-\omega_{m}|^{-1/4}$
and $\Lambda_{v}=\Lambda_{h,nm}=Z_{h,nm}^{-1}=1+\Lambda_{h}^{(0)}|\omega_{n}-\omega_{m}|^{-1/2}$
on the hot lines, but $\Lambda_{v}=\Lambda_{c,nm}$ on the cold parts
of the Fermi surface. The parameters $\Lambda_{c}^{(0)},\Lambda_{h}^{(0)}$
will be considered as tuning parameters controlling the strength of
the fluctuations.

To assess the impact of energy fluctuations on the $T_{c}$ for singlet
pairing of the hot quasiparticles, we tune the hot vertex pre-factor
$\Lambda_{h}^{(0)}$, a measure of the strength of hot pairing, from
$\Lambda_{h}^{(0)}\approx0.15$ up to $0.24$. These particular values are chosen because for $\Lambda_h^{(0)} < 0.15$, $2\pi T_c$ is above the energy cutoff of the energy fluctuation, whereas for $\Lambda_h^{(0)} > 0.24$, $T_c$ is below our numerical precision. Note that, because
$\Lambda_{h}^{(0)}$ only affects $\Lambda_{v}$, and because the
contribution to the pairing interaction arising from the energy fluctuations
has an overall $\Lambda_{v}$ pre-factor (see the $\alpha_{E}$ term
in Eq.~(\ref{V_general})), by changing $\Lambda_{h}^{(0)}$ we are
effectively changing the relative strength of the energy fluctuations
over the spin fluctuations. The strength of the cold vertex is kept
fixed as $\Lambda_{c}^{(0)}=0.5$. In addition, the AFM vector $\mathbf{Q}=\sqrt{2}k_{f}$
and thus $\theta_{0}=\pi/4$.

\begin{figure}[H]
\centering 
\includegraphics[width=0.42\textwidth]{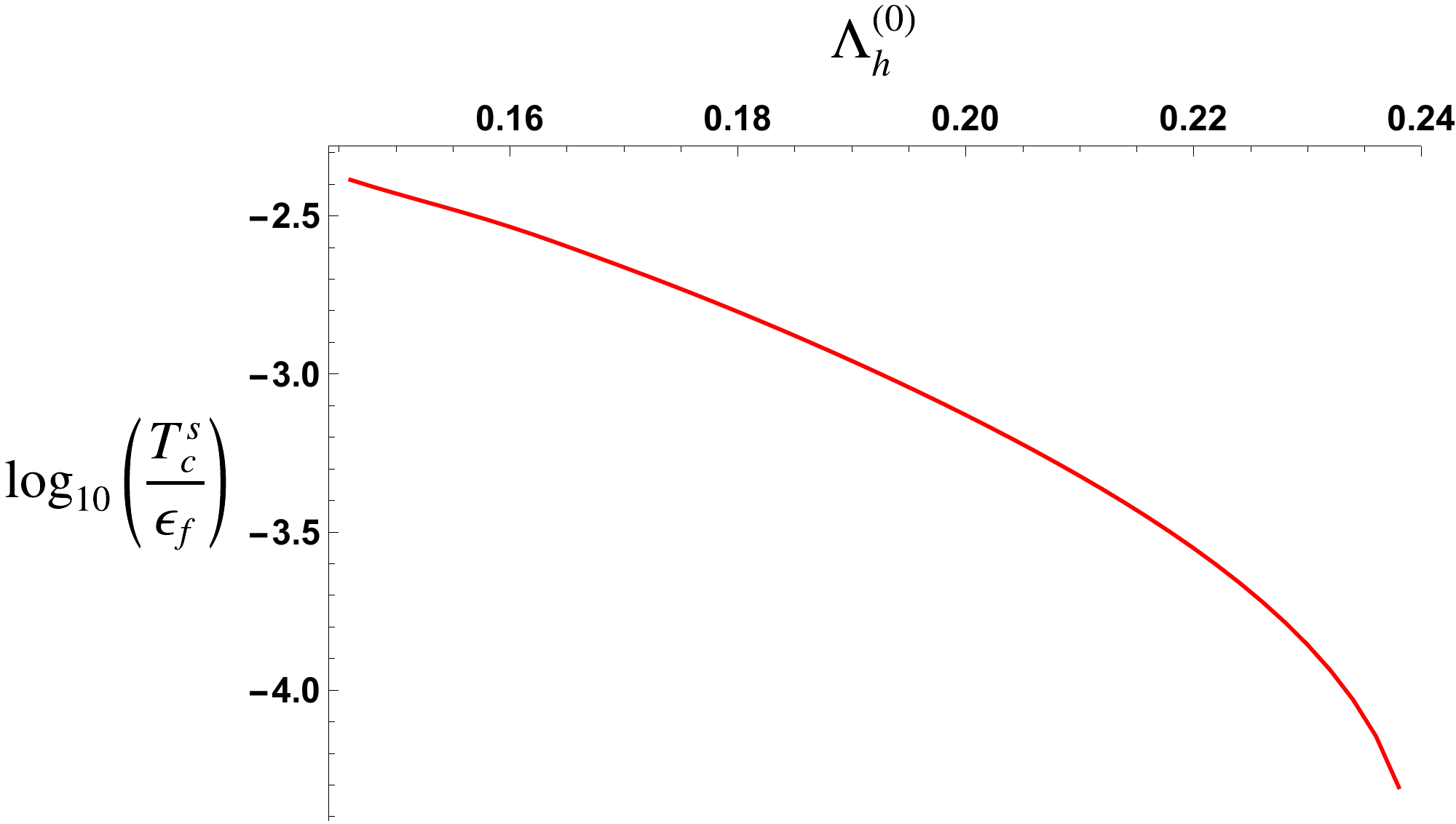} \caption{Suppression of $T_{c}$ by energy fluctuations. The pairing channel
considered here is the singlet channel, promoted by the exchange of
spin fluctuations between hot quasiparticles. $\Lambda_{h}^{(0)}$
denotes the strength of the energy-fluctuation vertex for the hot
quasiparticles. Here, $\epsilon_{F}$ is the Fermi energy.}
\label{Tcaplot}
\end{figure}

The resulting $T_{c}$ is plotted in Fig.~\ref{Tcaplot}. When the
energy fluctuations contribution is weaker ($\Lambda_{h}^{(0)}=0.15$),
a non-zero $T_{c}$ of order $0.004\epsilon_{F}$ is found at the
QCP ($r=0$). However, when the energy fluctuations contribution becomes
stronger, $T_{c}$ suffers a substantial suppression. This is in agreement
with experiments in YbRh$_{2}$Si$_{2}$, where superconductivity
appears to be absent in the expected temperature range of several
hundreds of mK. Another compound for which energy fluctuations are
thought to exist is CeCu$_{6-x}$Au$_{x}$ at $x\approx0.1$, where
again superconductivity has not been observed. In the latter, two-dimensional
anti-ferromagnetic spin fluctuations are thought to dominate and a
model calculation analogous to the one presented above applies. It
is also interesting to study $T_{c}$ without the contribution of
the energy fluctuations, i.e.~$Z_{h}=1$, which removes the $T_{c}$
suppression arising from the energy fluctuations from Eq.~(\ref{Eq:HotGapEqn}).
In this case, we found that $T_{c}/\epsilon_{f}\approx0.024$, or
$T_{c}\approx0.24$ K, suggesting that a strong suppression of $T_{c}$
by energy fluctuations is present in these compounds.

\subsection{Cold quasiparticles}
As discussed in Eq.\ (\ref{trip}), a triplet pairing interaction
is also generated by the exchange of energy fluctuations. We assume
$p$-wave symmetry as discussed above and consider the gap function
of the form $\Phi(\mathbf{k},i\omega_{n})=\Delta_{n}^{t}\cos\theta$,
where $\theta$ is the angle between $\mathbf{k}$ and the $z$-axis.
The linearized gap equation becomes

\begin{eqnarray}
\Delta_{n}^{t}\cos\theta_{k} & = & T\sum_{\omega_{m};\mathbf{p}}\frac{4N_{0}^{3}h_{nm}\alpha_{E,nm}^{2}\Lambda_{Q,nm}|\omega_{nm}|^{3/2}}{r+(\mathbf{k}-\mathbf{p})^{2}+\Lambda_{Q,nm}^{2}|\omega_{nm}|}\\
 &  & \times\frac{\Delta_{m}^{t}\cos\theta_{p}}{\omega_{m}^{2}Z_{m}^{-2}+\epsilon_{\mathbf{p}}^{2}},
\end{eqnarray}
In contrast to the case of hot quasiparticles, the vertex function for the cold quasiparticles is $\lambda_v = \Lambda_c$, resulting in the coupling constant $\alpha_{E} \approx \Lambda_c \big( \Lambda_Q / N_0 \big)^2$.
Performing the momentum
integral in a similar way as in the singlet pairing case and again
defining $f_{m}^{t}=\Delta_{m}^{t}Z_{m}/|\omega_{m}|$, the following
eigenvalue problem in Matsubara frequency space is found:
\begin{equation}
\sum_{\omega_{m}}K_{n,m}^{t}f_{m}^{t}=0\label{coldgap}
\end{equation}

The kernel is given by
\begin{align}
K_{n\neq m}^{t} & =\frac{1}{2}W_{nm}^{t},\nonumber \\
K_{n,n}^{t} & =-(2n+1)Z_{n}^{-1}+\frac{1}{2}(K_{n,n-1}^{t}+K_{n-1,n}^{t})
\end{align}
where
\begin{equation}
W_{nm}^{t}=4h_{nm}\Lambda_{c,nm}^{7}|\Omega_{nm}|^{3/2}I_{nm}^{t}
\end{equation}
and
\begin{eqnarray}
I_{nm}^{t} & = & (2+B_{nm}^{t})\ln(1+4/B_{nm}^{t})-4,\nonumber \\
B_{nm}^{t} & = & \Lambda_{c,nm}^{2}|\Omega_{nm}|
\end{eqnarray}

Again, $h_{nm}=h(\omega_{n}-\omega_{m})$\ is the soft cutoff function
introduced above.

\begin{figure}[H]
{\centering 
\includegraphics[width=0.42\textwidth]{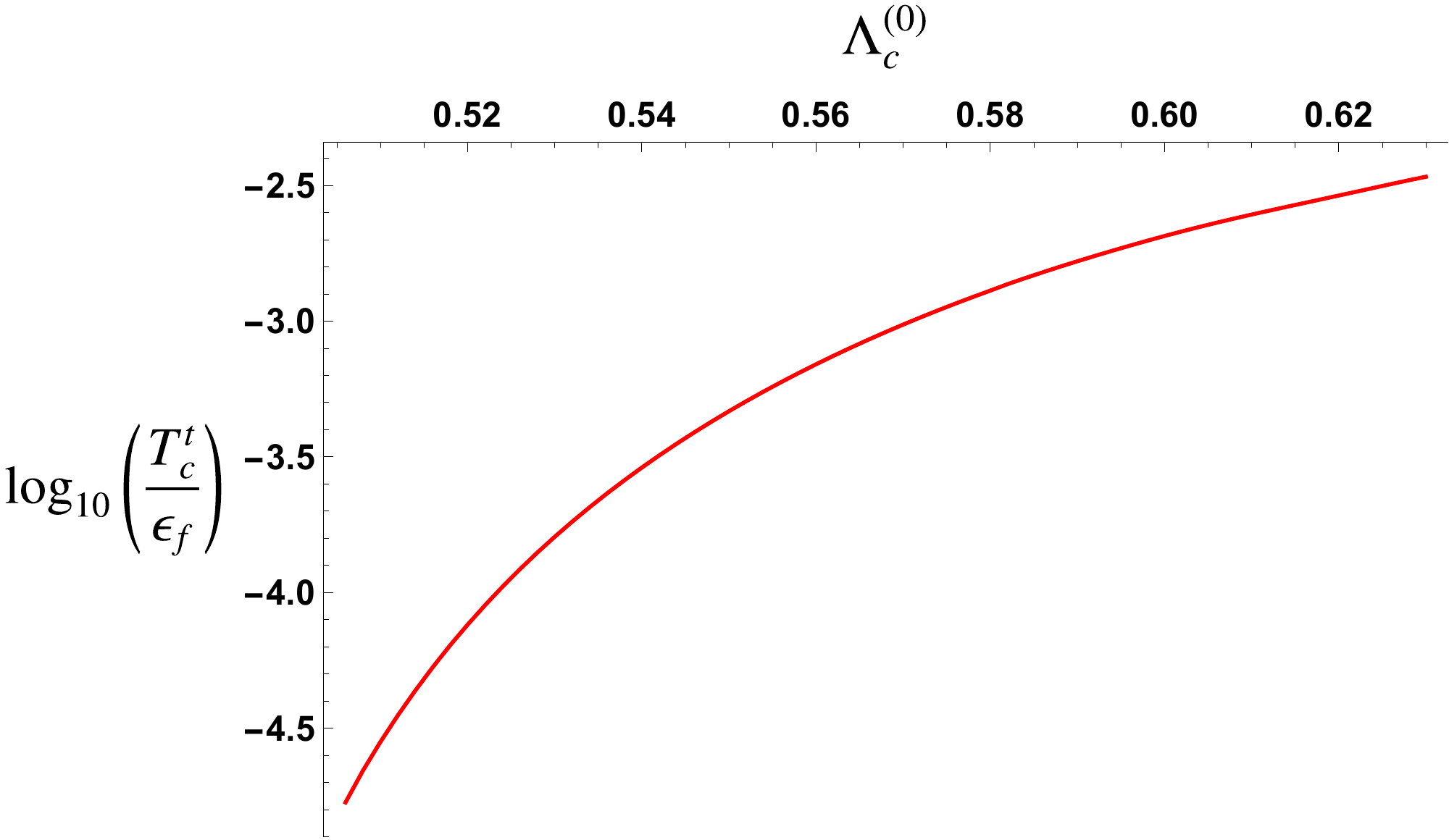}} \caption{Superconducting transition temperature $T_{c}^{t}$ in the triplet
pairing channel. This pairing is mediated by the exchange of energy
fluctuations by cold quasi-particles. $\Lambda_{c}^{(0)}$ denotes
the strength of the energy fluctuation vertex for cold quasiparticles.}
\label{triplet}
\end{figure}

The $T_{c}$ values for triplet pairing obtained by numerical solution
of Eq.~(\ref{coldgap}) are shown in Fig.~\ref{triplet} as function
of the bare vertex strength $\Lambda_{c}^{(0)}$. A strong dependence
on $\Lambda_{c}^{(0)}$ is found. In particular, for the value $\Lambda_{c}^{(0)}=0.5$
that we chose for the singlet pairing solution, we find $T_{c}/\epsilon_{F}\approx1.5\times10^{-5}$,
corresponding to $T_{c}\approx0.15$mK, as compared to the singlet
pairing $T_{c}\approx0.24$K found in the absence of energy fluctuations.
It remains to be seen whether the superconducting phase observed \cite{schu}
in YbRh$_{2}$Si$_{2}$ at mili-Kelvin temperatures is of spin-triplet
symmetry, which our calculations suggest to be a possibility.

\section{Conclusions}

Motivated by recent experimental evidence \cite{schu} for superconductivity
at extremely low temperature in \yrs, we have used the recently-developed
theory of critical quasiparticles \cite{qp1,WSA} to discuss the superconductivity
generated by pairing mediated by critical fluctuations in the neighborhood
of an antiferromagnetic quantum critical point, which is often present
in the phase diagram of heavy-fermion compounds. 
In these materials, critical antiferromagnetic spin fluctuations
are dominant and are responsible for many of the observed properties
near the critical region. Since these fluctuations have a non-zero
wavevector ${\bf Q}$, usually of order $k_{F}$, they divide the
Fermi surface into hot regions, which are connected by ${\bf Q}$,
and cold regions, which are not. This usually leads to unconventional
pairing (\textit{e.g.} $d$-wave) of hot quasiparticles as is seen
in cuprates and some heavy-fermion superconductors.

However, as emphasized in Refs.\ \onlinecite{exp2, WSA,hart}, composite
critical spin fluctuations induce energy fluctuations at small momentum,
leading to a diverging quasiparticle effective mass over the {\em
whole} Fermi surface. This contibution is essential to achieve the
excellent agreement between the critical quasiparticle theory with
the experimental results for thermodynamic and transport quantities
on \cca and \yrs. In this paper, we studied the impact of these
energy fluctuations on the pairing channel by employing an Eliashberg-like
approach. Our main results are that, while the exchange of energy
fluctuations suppresses the $d$-wave $T_{c}$ of hot quasiparticles,
they can at the same time mediate spin-triplet (e.g. $p$-wave) superconductivity
of cold quasiparticles, a possibility that can be probed experimentally,
for example using NMR.

\section{Acknowledgments}

JK, RMF, and PW are grateful to have had the pleasure of collaborating
in this project with EA, who passed away during the final stages of
preparation of this manuscript. The authors acknowledge fruitful discussions
with A. Chubukov, and J. Schmalian. JK was supported by the
National High Magnetic Field Laboratory through NSF Grant No.~DMR-1157490
and the State of Florida. RMF was supported by the U.S. Department
of Energy, Office of Science, Basic Energy Sciences, under Award No.~DE-SC0012336. Part of this work was developed when the authors attended
research programs at the Aspen Center for Physics (RMF, EA, and PW),
which is supported by the National Science Foundation under Grant
No.~PHY-1066293, and at KITP (JK, RMF and EA), which is supported by the
National Science Foundation under Grant No.~PHY17-48958.

\end{document}